\documentclass[useAMS,usenatbib]{mn2e}

\usepackage{natbib}
\usepackage{graphicx}
\usepackage{amssymb}
\usepackage{color}
\usepackage[dvipsnames]{xcolor}
\usepackage{caption}
\usepackage{lipsum,graphicx,multicol}
\usepackage{float}
\usepackage[fleqn]{amsmath}

\usepackage{subcaption}

\title[Early supercritical dusty BH growth]{Supercritical dusty BH growth in the early Universe}

\author[ ]
{W. Ishibashi$^{1}$\thanks{E-mail: wako.ishibashi@physik.uzh.ch} 
\vspace{0.05cm}
\footnotemark[0]\\
$^{1}$Physik-Institut, Universitat Zurich, Winterthurerstrasse 190, 8057 Zurich, Switzerland 
}

\begin{document}

\pdfminorversion=4

\date{Accepted ? Received ?; in original form ? }

\pagerange{\pageref{firstpage}--\pageref{lastpage}} \pubyear{2012}

\maketitle

\label{firstpage}

\begin{abstract} 
Supermassive black holes (with $\mathrm{M_{BH} \sim 10^9 M_{\odot}}$) are observed in the first Gyr of the Universe, and their host galaxies are found to contain unexpectedly large amounts of dust and metals. In light of the two empirical facts, we explore the possibility of supercritical accretion and early black hole growth occurring in dusty environments. We generalise the concept of photon trapping to the case of dusty gas and analyse the physical conditions leading to `dust photon trapping'. Considering the parameter space dependence, we obtain that the dust photon trapping regime can be more easily realised for larger black hole masses, higher ambient gas densities, and lower gas temperatures. The trapping of photons within the accretion flow implies obscured active galactic nuclei (AGNs), while it may allow a rapid black hole mass build-up at early times. We discuss the potential role of such dust photon trapping in the supercritical growth of massive black holes in the early Universe. 
\end{abstract} 



\section{Introduction}

Observational surveys over the past years have uncovered the existence of supermassive black holes with masses of $\mathrm{M_{BH} \sim 10^9 M_{\odot}}$ in the first Gyr of the Universe. Today, a few hundred quasars are known at redshifts $z \gtrsim 6$, which may only represent the tip of the iceberg \citep[][and references therein]{Inayoshi_et_2020}. In addition to the two quasars known at $z \sim 7.5$ \citep{Banados_et_2018, Yang_et_2020}, the latest discovery of the most distant quasar at $z = 7.642$, with a black hole (BH) mass of $\mathrm{M_{BH} = (1.6 \pm 0.4) \times 10^9 M_{\odot}}$, is reported in \citet{Wang_et_2021}. The estimated Eddington ratios are high ($L/L_E \sim 0.7-1$), suggesting that these massive luminous quasars represent extreme objects accreting close to the Eddington limit at early times. 

In contrast, the majority of the accreting black hole population in the early Universe are likely to have lower luminosities, either due to smaller masses or lower accretion rates. Such low-luminosity quasars are now starting to be detected in the Subaru High-z Exploration of Low-Luminosity Quasars (SHELLQs) survey \citep[][]{Matsuoka_et_2019}. In addition, many accreting black holes may be obscured by gas and dust, hence still awaiting to be discovered. 

From a theoretical perspective, the mere existence of supermassive black holes (with $\mathrm{M_{BH} \sim 10^9 M_{\odot}}$) at the highest redshifts ($z > 6$) poses severe challenges to black hole growth models \citep[][and references therein]{Woods_et_2019}. 
In the case of light seeds ($\mathrm{M_0 \sim 10^2 - 10^3 M_{\odot}}$, such as remnants of Population III stars), a continuous growth at the Eddington rate and/or several super-Eddington accretion episodes are required. Even in the case of heavy seeds ($\mathrm{M_{BH} \sim 10^4 - 10^5 M_{\odot}}$, such as direct collapse black holes or DCBH), a rapid and efficient growth is still required in order to account for the billion-solar masses observed at $z \gtrsim 6$. In any case, some form of supercritical accretion may be required at very high redshifts \citep[e.g.][]{Volonteri_Rees_2005}. 

The Bondi accretion rate gives an order of magnitude estimate of how rapidly infalling gas can be supplied from large scales to the central BH, in the case of spherical symmetry and without any feedback. But, as a result of high accretion rates, a huge amount of radiation is released into the surrounding environment. The resulting radiative feedback can suppress the gas inflow, and the actual accretion rate may be limited to the Eddington rate defined by the balance between radiation pressure and gravity. 

Numerical simulations suggest that the mass accretion rate can also be considerably reduced due to photoionisation feedback 
\citep{Milosavljevic_et_2009, Park_Ricotti_2011, Park_Ricotti_2012}. In this picture, photons heat and ionise the surrounding gas, creating an expanding hot bubble (HII region) opposing the inflowing gas. 2D hydrodynamic simulations show that the accretion follows an oscillatory pattern with a quasi-periodic trend, due to both photoionisation heating and radiation pressure, with an average accretion rate of $\sim 0.2 \%$ of the Bondi rate \citep{Milosavljevic_et_2009}. 

\citet{Park_Ricotti_2011} perform a suite of 1D and 2D hydrodynamic simulations with radiative transfer in spherical symmetry, and obtain an oscillatory behaviour for the accretion rate that is attributed to UV and X-ray photoheating. In this case, the mean accretion rate is $\sim 3 \%$ of the Bondi rate, with a steep dependence on the ambient gas temperature. A qualitatively similar picture is found in subsequent simulations, including the effects of angular momentum and radiation pressure, with a somewhat lower average accretion rate of $\sim 1\%$ of the Bondi rate \citep{Park_Ricotti_2012}. Overall, these numerical simulations indicate that the time-averaged accretion rate is mostly limited to a small fraction of the Bondi rate due to radiative feedback. 

On the other hand, photon trapping can be realised for sufficiently high accretion rates. In this case, the trapping of photons within the accretion flow reduces the impact of radiative feedback, thus potentially allowing supercritical accretion \citep[][]{Wyithe_Loeb_2012}. In fact, if the emergent luminosity is limited by photon trapping ($L \lesssim L_E$), the central radiation can not much affect the gas dynamics on larger scales. Several episodes of super-Eddington accretion may then account for the rapid growth of massive black holes at high redshifts \citep{Volonteri_et_2015}. In this context, \citet{Pacucci_et_2017} discuss the most favourable conditions leading to effective BH growth at early times. 

Recent radiation hydrodynamic (RHD) simulations suggest that a transition to a `hyper-Eddington' regime can occur at very high accretion rates in high-density environments. The required condition is that the size of the ionised HII region is smaller than the Bondi radius ($\mathrm{R_\mathrm{HII} < R_B}$, where $\mathrm{R_B = G M_{BH}/c_s^2}$ with the gas sound speed $c_s$); and 1D RHD simulations indicate that the resulting accretion rate can reach the Bondi value \citep{Inayoshi_et_2016}. This result seems to hold even for super-Eddington luminosities \citep{Sakurai_et_2016}. 

Although most of the above mentioned numerical simulations consider primordial gas accretion, recent observations indicate the existence of metal-enriched gas at very high redshifts. Intriguingly, the host galaxies of the first quasars are found to contain large amounts of dust and metals. ALMA observations of a sample of $z > 6$ quasars indicate typical dust masses in the range $\mathrm{M_d \sim (10^7 - 10^9) M_{\odot}}$ \citep{Venemans_et_2018}. The common presence of dust grains at such high redshifts suggests early enrichment and a rapid build-up of the dust reservoir. Therefore the early growth of massive black holes is likely to occur in dusty environments. 

In light of the latest observational results, namely the simultaneous detection of supermassive black holes and dust reservoirs at $z > 6$, here we explore the possibility of supercritical accretion occurring in the dusty environment of the first quasars. We generalise the concept of photon trapping to the case of dusty gas, and analyse whether dust photon trapping can be actually realised in the early Universe.


\section{Photon trapping in dusty gas}
\label{Subsection_photon_trapping}

Photon trapping occurs when the infall velocity of the accreting gas ($v_r$) exceeds the outward diffusion speed of the radiation ($c/\tau$) \citep{Begelman_1979}. The optical depth is given by $\tau = \kappa_e \rho r$, where $\kappa_\textrm{e}$ is the electron scattering opacity, and $\rho$ is the gas density related to the accretion rate by $\dot{M} = 4 \pi r^2 \rho v_r$. 
The condition for the trapping of photons ($v_r > c/\tau$) is satisfied for radii smaller than a critical radius
\begin{equation}
R_\mathrm{tr,e} = \frac{\kappa_\mathrm{e} \dot{M}}{4 \pi c} . 
\label{Eq_R_tre}
\end{equation}
This is the photon trapping radius, within which radiation is advected inwards. The standard photon trapping radius (equation \ref{Eq_R_tre}) is determined by electron scattering ($\kappa_e = \sigma_T/m_p\cong 0.4 \, \rm{{cm^2} g^{-1}}$, where $\sigma_T$ is the Thomson cross section) and only scales with the physical accretion rate $\dot{M}$. We note that the minimal condition for photon trapping is that the critical radius in equation \ref{Eq_R_tre} lies outside the Schwarzschild radius: $\mathrm{R_{tr,e} > R_S}$ (where $\mathrm{R_S = \frac{2 G M_{BH}}{c^2}}$), which implies a minimal accretion rate $\mathrm{\dot{M} > \dot{M}_e = \frac{8 \pi G M_{BH}}{\kappa_e c}}$. This can be compared to the Eddington accretion rate $\mathrm{\dot{M}_E = L_E/(\epsilon c^2)}$, where $\epsilon$ is the radiative efficiency. The resulting ratio is $\mathrm{\frac{\dot{M}_e}{\dot{M}_E} \sim 0.2 \left( \frac{\epsilon}{0.1} \right)}$, suggesting that the minimal condition for photon trapping due to electron scattering can be easily satisfied, even for sub-Eddington accretion rates. 

Electron scattering opacity dominates in the innermost region close to the central BH, but dust opacity becomes more important on larger scales. While the inner zone is dust-free, dust grains can survive at larger radii where the temperature drops below the dust sublimation temperature. The corresponding dust sublimation radius is estimated by considering the energy balance between the absorbed and reradiated fluxes \citep[e.g.][]{Murray_et_2005} and is given by $R_\textrm{sub} = \sqrt{\frac{L}{4 \pi \sigma_\textrm{sb} T_\textrm{sub}^4}}$, where $L = f_\mathrm{E} L_\mathrm{E}$ is the luminosity expressed in units of the Eddington luminosity, $\sigma_\mathrm{sb}$ is the Stefan-Boltzmann constant, and $T_\mathrm{sub}$ is the sublimation temperature of dust grains.

By analogy with equation (\ref{Eq_R_tre}), we define a photon trapping radius for dusty gas, where the opacity is dominated by dust absorption ($\kappa_d = \sigma_d/m_p$, where $\sigma_d$ is the dust absorption cross section)
\begin{equation}
R_\mathrm{tr,d} = \frac{\kappa_d \dot{M}}{4 \pi c}  . 
\label{Eq_R_trd}
\end{equation}
Since the radiation-matter coupling is enhanced in the presence of dust grains (with a typical ratio of $\kappa_d/\kappa_e \gtrsim 10^3$ in the ultraviolet band), the dust photon trapping radius is correspondingly larger: $R_\mathrm{tr,d} = (\kappa_d/\kappa_e) R_\mathrm{tr,e}$. 

Dust photon trapping (DPT) can only be achieved if the actual trapping radius (equation \ref{Eq_R_trd}) exceeds the sublimation radius of dust grains 
\begin{equation}
R_\mathrm{tr,d} > R_\mathrm{sub} . 
\end{equation}
The above condition implies a minimal accretion rate for dusty gas: $\dot{M} > \dot{M}_\mathrm{d}$. 
The critical accretion rate for DPT is given by
\begin{align}
&\dot{M}_\mathrm{d} 
= \frac{4 \pi c}{\kappa_\mathrm{d}} \sqrt{ \frac{G c f_\mathrm{E} M_\mathrm{BH}}{\kappa_\mathrm{e} \sigma_\mathrm{sb} T_\mathrm{sub}^4} }   \nonumber \\
&\cong 0.3 \frac{M_{\odot}}{\mathrm{yr}} \, \left(\frac{\kappa_\mathrm{d}}{10^3 \mathrm{cm^2/g}}\right)^{-1} \left( \frac{f_\mathrm{E}}{1} \right)^\frac{1}{2}  
\left( \frac{T_\mathrm{sub}}{1500 \mathrm{K}} \right)^{-2} \left( \frac{M_\mathrm{BH}}{10^5 M_{\odot}} \right)^\frac{1}{2} 
\label{Eq_Mdotd}
\end{align}  
In the following, we analyse under which physical conditions this critical DPT accretion rate can be achieved, and whether dust photon trapping may be actually realised. 

Here we implicitly consider a quasi-spherical accretion assuming that the accreting gas has negligible angular momentum. The importance of low-angular momentum gas in promoting supercritical accretion has been emphasised by \citet{Volonteri_et_2015}.
We also note that spherical symmetry is adopted in 1D RHD simulations of hyper-Eddington BH growth \citep[e.g.][]{Inayoshi_et_2016}. 

In more realistic situations, the infalling gas has non-zero angular momentum, leading to the formation of a compact accretion disc. 
By conservation of specific angular momentum, the disc size may be expressed as $\mathrm{R_\textrm{disc} \sim \lambda^2 R_{\sigma} \sim \lambda^2 (G M_\mathrm{BH}/\sigma^2)}$, where $\lambda \leq 1$ is the fraction of preserved angular momentum \citep{Volonteri_et_2015}. The whole accretion disc radiation can be efficiently trapped if the disc radius is smaller than the photon trapping radius ($R_\mathrm{disc} < R_\mathrm{tr}$). We note that this condition is more easily satisfied for dust photon trapping than for electron scattering (since $\mathrm{R_{tr,d} \gg R_{tr,e}}$). In fact, for a typical value of $\lambda \sim 0.05$, the condition for efficient disc radiation trapping ($\mathrm{R_\mathrm{disc} < R_{tr,d}}$) is naturally satisfied for DPT accretion rates. Evidently, a lower $\lambda$ would imply a smaller accretion disc, further facilitating the trapping condition. A particular solution of supercritical accretion flows is the slim accretion disc model \citep[][]{Abramowicz_et_1988, Czerny_2019}. In the context of disc models, the trapping radius may be adjusted by a factor $h = H/R$, corresponding to the disc aspect ratio, which may be of order unity for geometrically thick discs puffed up by trapped radiation \citep{Ohsuga_et_2005}. 


\section{Dust photon trapping: parameter space analysis}
\label{Section_DPT_parameterspace}

\begin{figure}
\begin{center}
\includegraphics[angle=0,width=0.45\textwidth]{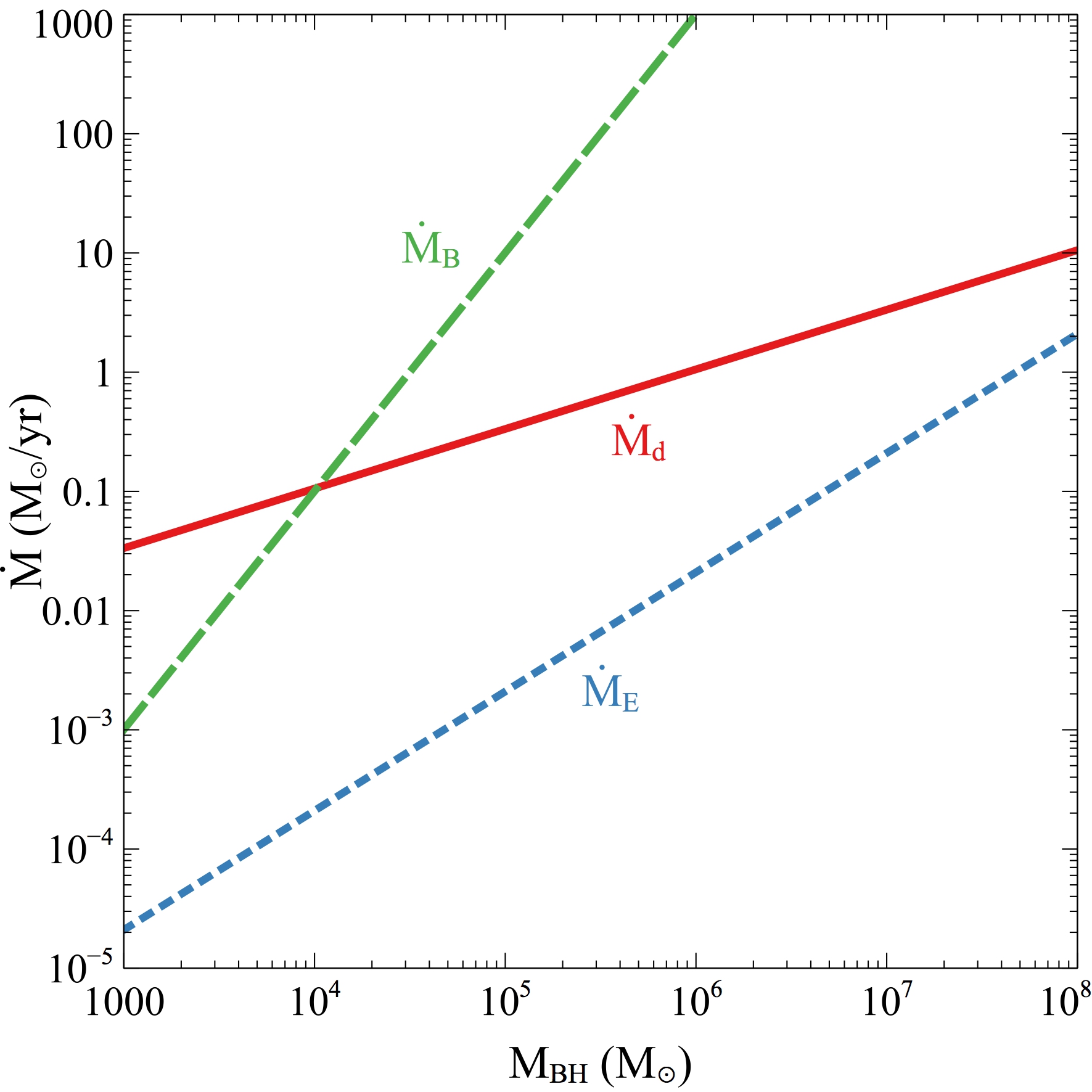}
\caption{Characteristic accretion rates as a function of black hole mass (with fiducial parameters, see the main text): Eddington rate (blue dotted), DPT rate (red solid), and Bondi rate (green dashed). 
}
\label{Fig_dotM_MBH}
\end{center}
\end{figure} 

\begin{figure*}
\begin{multicols}{2}
    \includegraphics[width=0.9\linewidth]{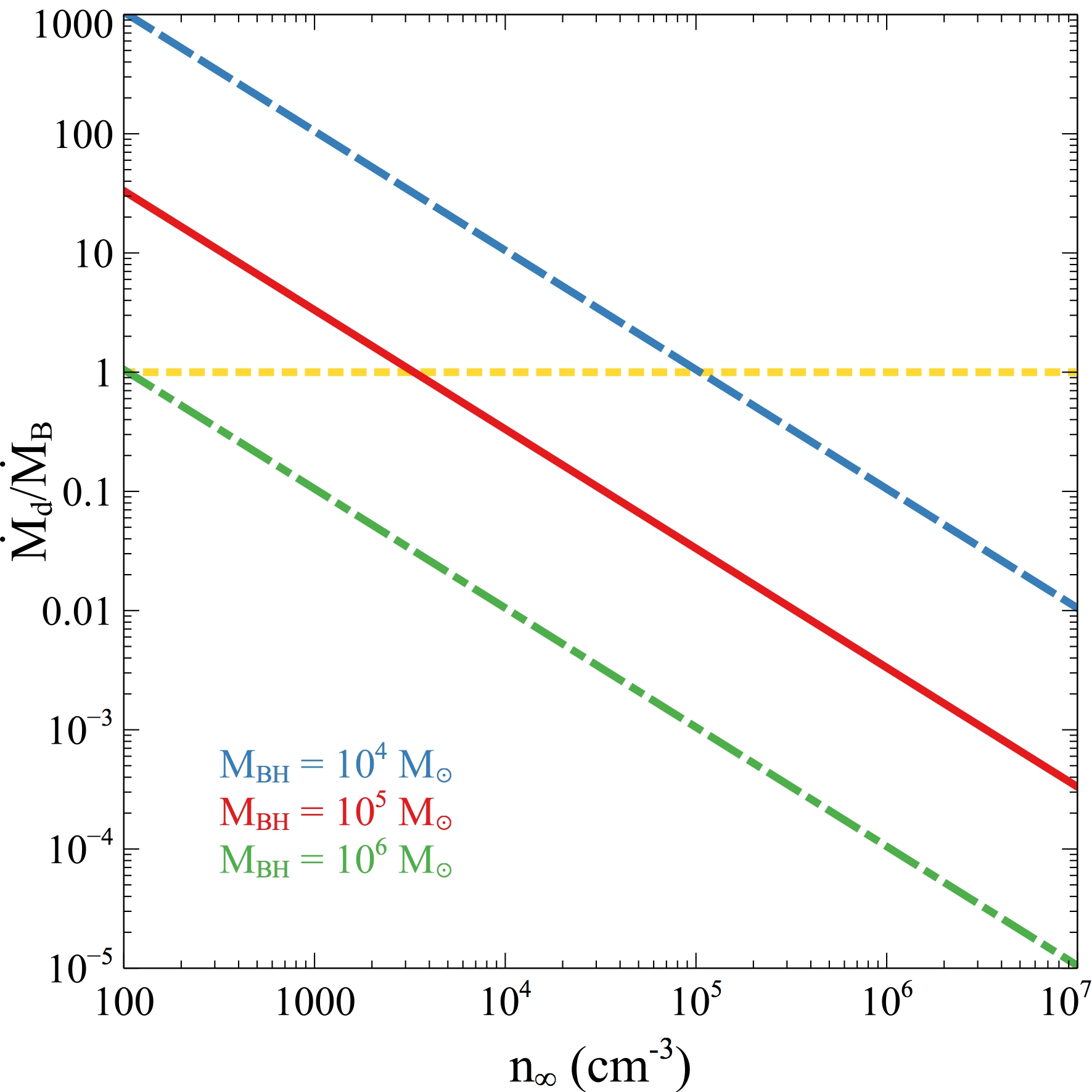}\par
    \includegraphics[width=0.9\linewidth]{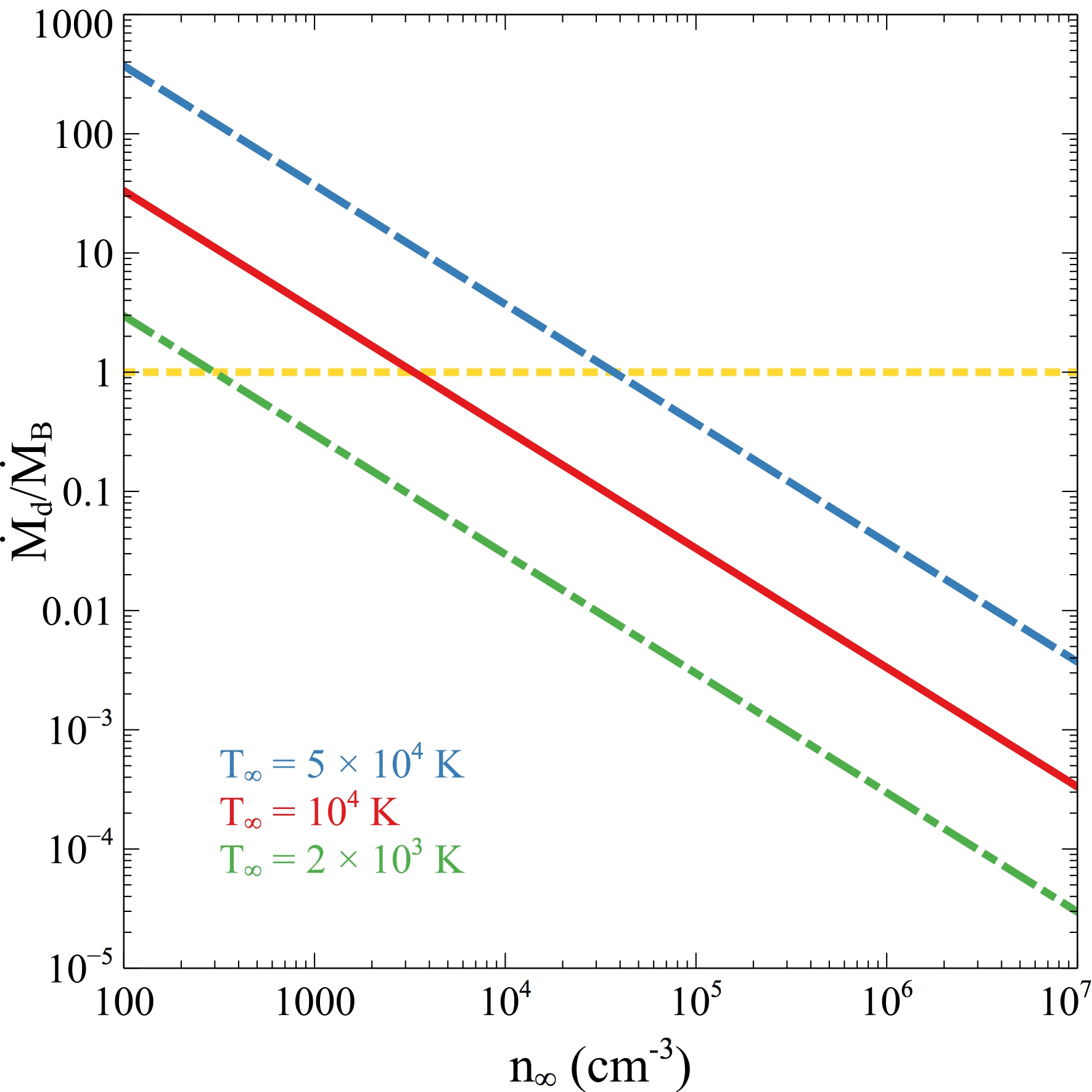}\par
    \end{multicols}
\caption{DPT accretion rate normalised by the Bondi rate versus ambient gas density. DPT is realised for $\mathrm{\dot{M}_d/\dot{M}_B < 1}$, i.e. when the ratio falls below unity (marked by the horizontal yellow dotted line). Left panel: variations in the black hole mass; right panel: variations in the ambient gas temperature. 
}
\label{Fig_MdotB_ninf}
\end{figure*}

The critical accretion rate for dust photon trapping (equation \ref{Eq_Mdotd}) can be compared with the Eddington accretion rate, and the ratio between the two is given by
\begin{align}
&\frac{\dot{M}_\mathrm{d}}{\dot{M}_\mathrm{E}} = \epsilon c^2 \frac{\kappa_\mathrm{e}}{\kappa_\mathrm{d}} \sqrt{ \frac{c f_\mathrm{E}}{\kappa_\mathrm{e} \sigma_\mathrm{sb} T_\mathrm{sub}^4 G M_\mathrm{BH}} } \\ \nonumber
&\cong 160 \left( \frac{\epsilon}{0.1} \right) \left( \frac{f_\mathrm{E}}{1} \right)^\frac{1}{2} \left( \frac{\kappa_\mathrm{d}}{10^3 \mathrm{cm^2/g}} \right)^{-1}  \left( \frac{T_\mathrm{sub}}{1500 \mathrm{K}} \right)^{-2} \left( \frac{M_\mathrm{BH}}{10^5 M_{\odot}} \right)^{-\frac{1}{2}} \nonumber
\label{Eq_Md_ME}
\end{align} 

We see that the condition for dust photon trapping requires super-Eddington accretion for typical parameters. 
Moreover, this implies that DPT accretion naturally includes photon trapping by electron scattering, while the converse is not true. 

In order to realise dust photon trapping, the critical accretion rate must be smaller than the Bondi accretion rate. 
The Bondi rate is given by $\dot{M}_B = (4 \pi m_p G^2 M_\mathrm{BH}^2 n_{\infty})/c_s^3$ , where $c_s = \sqrt{ \frac{k_B T_{\infty}}{\mu m_p}}$ is the sound speed of the gas, with $T_{\infty}$ and $n_{\infty}$ being the gas temperature and gas number density, respectively, and $\mu$ is the mean molecular weight ($\mu = 1.22$). 
The ratio between the DPT and Bondi accretion rates is given by 
\begin{align}
\frac{\dot{M}_{d}}{\dot{M}_B} 
&= \frac{1}{m_p \kappa_d n_{\infty}} \left( \frac{f_E}{\kappa_e \sigma_{SB} T_{sub}^4} \right)^{1/2} \left( \frac{c k_B T_{\infty}}{ \mu m_p G M_{BH}} \right)^{3/2}  \nonumber \\
&\cong 0.03 \left( \frac{f_E}{1} \right)^{1/2} \left( \mathrm{\frac{\kappa_d}{10^3 cm^2/g}} \right)^{-1} \left( \mathrm{\frac{T_{sub}}{1500 K}} \right)^{-2} \nonumber \\ 
\quad \quad \quad &\left( \mathrm{\frac{T_{\infty}}{10^4 K}} \right)^{3/2} \left( \mathrm{\frac{n_{\infty}}{10^5 cm^{-3}}} \right)^{-1} \left( \frac{M_{BH}}{10^5 M_{\odot}} \right)^{-3/2}
\end{align}
We see that the ratio is smaller than unity ($\dot{M}_\textrm{d}/\dot{M}_\textrm{B} < 1$) for typical parameters, suggesting that DPT can be potentially realised. 
The reference values for the ambient gas density and temperature are set to $n_{\infty} = 10^5 \mathrm{cm}^{-3}$ and $T_{\infty} = 10^4 \, \mathrm{K}$, as adopted in numerical simulations \citep{Inayoshi_et_2016, Takeo_et_2018}. 

As mentioned in the Introduction, the mean accretion rate can be much lower than the Bondi rate due to photoionisation feedback, typically of the order of $\sim 1\%$ \citep{Park_Ricotti_2011, Park_Ricotti_2012}. In contrast, in sufficiently dense environments characterised by efficient photon trapping, the accretion rate can approach the Bondi value \citep{Inayoshi_et_2016}. Hence Bondi accretion rates may be achieved under certain physical conditions, without being much hampered by radiative feedback. Here we consider the simple case of gas accretion onto a static BH and consider Bondi accretion as the limiting case (for BHs moving through the surrounding medium, see e.g. \citet{Toyouchi_et_2020, Sugimura_Ricotti_2020}). We note in passing that the DPT accretion rate is a few percent of the Bondi rate for fiducial parameters (comparable to the radiation feedback-reduced Bondi rate). 

In Figure $\ref{Fig_dotM_MBH}$, we compare the three critical accretion rates (Eddington rate, Bondi rate, and DPT rate) as a function of the black hole mass. We note their different scalings with BH mass: $\dot{M}_\textrm{E} \propto M_\textrm{BH}$, $\dot{M}_\textrm{B} \propto M_\textrm{BH}^2$, and $\dot{M}_\textrm{d} \propto \sqrt{M_\textrm{BH}}$, respectively. 
We observe that the critical accretion rate for dust photon trapping is always larger than the Eddington rate, thus implying super-Eddington accretion. As the black hole mass increases, the difference between the two becomes less prominent (due to the shallower mass-dependence of the DPT rate, $\dot{M}_\textrm{d} \propto \sqrt{M_\textrm{BH}}$). The Bondi accretion rate exceeds the DPT rate above a certain critical mass ($\mathrm{M_\textrm{BH} \gtrsim 10^4 M_{\odot}}$); whereas for lower black hole masses, the DPT accretion rate is larger than the Bondi rate. Therefore, a certain minimal black hole mass is required in order to achieve dust photon trapping. 
The condition $\dot{M}_\textrm{d}/\dot{M}_\textrm{B} < 1$ implies a critical BH mass of 
\begin{align}
M_\mathrm{BH} &> \frac{c k_B T_{\infty}}{G \mu m_p} \left( \frac{1}{m_p \kappa_d n_{\infty}} \right)^{2/3} \left( \frac{f_E}{\kappa_e \sigma_{SB} T_{sub}^4} \right)^{1/3}  \nonumber \\
 &> 10^4 \, M_{\odot} \, \left( \frac{f_E}{1} \right)^{1/3} \left( \mathrm{\frac{\kappa_d}{10^3 cm^2/g}} \right)^{-2/3} \left( \mathrm{\frac{T_{sub}}{1500 K}} \right)^{-4/3} \nonumber \\& 
\quad \quad \quad \quad \quad \left( \mathrm{\frac{T_{\infty}}{10^4 K}} \right) \left( \mathrm{\frac{n_{\infty}}{10^5 cm^{-3}}} \right)^{-2/3} \, , 
\label{Eq_MBH_d}
\end{align} 
which corresponds to the critical mass observed in Figure $\ref{Fig_dotM_MBH}$. 
We also note that the difference between the DPT and Bondi accretion rates increases with increasing BH mass (due to the steeper mass-dependence of the Bondi rate $\mathrm{\dot{M}_B \propto M_{BH}^2}$), hence the DPT condition is more easily satisfied at larger masses. 

Equivalently, the DPT condition can also be expressed in terms of a critical ambient gas density as
\begin{align}
n_{\infty} > 3 \times 10^3 \mathrm{cm}^{-3} &\left( \frac{f_E}{1} \right)^{1/2}  \left( \mathrm{\frac{\kappa_d}{10^3 cm^2/g}} \right)^{-1} \left( \mathrm{\frac{T_{sub}}{1500K}} \right)^{-2} \nonumber \\
&\left( \mathrm{\frac{T_{\infty}}{10^4K}} \right)^{3/2} \left( \frac{M_{BH}}{10^5 M_{\odot}} \right)^{-3/2} \, . 
\label{Eq_ninf_d}
\end{align}  

In Figure \ref{Fig_MdotB_ninf} we plot the DPT accretion rate normalised by the Bondi rate ($\dot{M}_d/\dot{M}_B$) as a function of the ambient gas density. Dust photon trapping occurs when this ratio is smaller than unity. The left-hand panel of Figure \ref{Fig_MdotB_ninf} shows variations in the central BH mass. For the fiducial case with $\mathrm{M_{BH} = 10^5 M_{\odot}}$, the ratio becomes smaller than unity at gas densities larger than $\mathrm{n_{\infty} \gtrsim 3 \times 10^3 cm^{-3}}$ (cf. equation \ref{Eq_ninf_d}), where DPT can be realised. 
For a smaller BH mass ($\mathrm{M_{BH} = 10^4 M_{\odot}}$), the gas density must exceed $\mathrm{n_{\infty} \gtrsim 10^5 cm^{-3}}$ to achieve DPT; while for a larger BH mass ($\mathrm{M_{BH} = 10^6 M_{\odot}}$), DPT can already be reached at lower densities of $\mathrm{n_{\infty} \gtrsim 10^2 cm^{-3}}$. Thus a larger BH mass allows the system to enter the DPT regime for lower gas densities and vice-versa. In fact, the $\dot{M}_{d}/\dot{M}_B$ ratio decreases with increasing gas density, scaling as $\propto 1/n_{\infty}$. This implies that at high gas densities, the DPT condition $\dot{M}_\textrm{d}/\dot{M}_\textrm{B} < 1$ is more easily satisfied, hence dust photon trapping can be realised for lower-mass black holes. 
The right-hand panel of Figure \ref{Fig_MdotB_ninf} shows variations in the ambient gas temperature. In the fiducial case with $\mathrm{T_{\infty} = 10^4 K}$, the $\dot{M}_{d}/\dot{M}_B$ ratio becomes smaller than unity and DPT occurs for $\mathrm{n_{\infty} \gtrsim 3 \times 10^3 cm^{-3}}$. 
We see that for a lower gas temperature, DPT can be reached at lower gas densities, while higher densities are required for a higher ambient gas temperature. 

\begin{figure*}
\begin{multicols}{2}
    \includegraphics[width=0.9\linewidth]{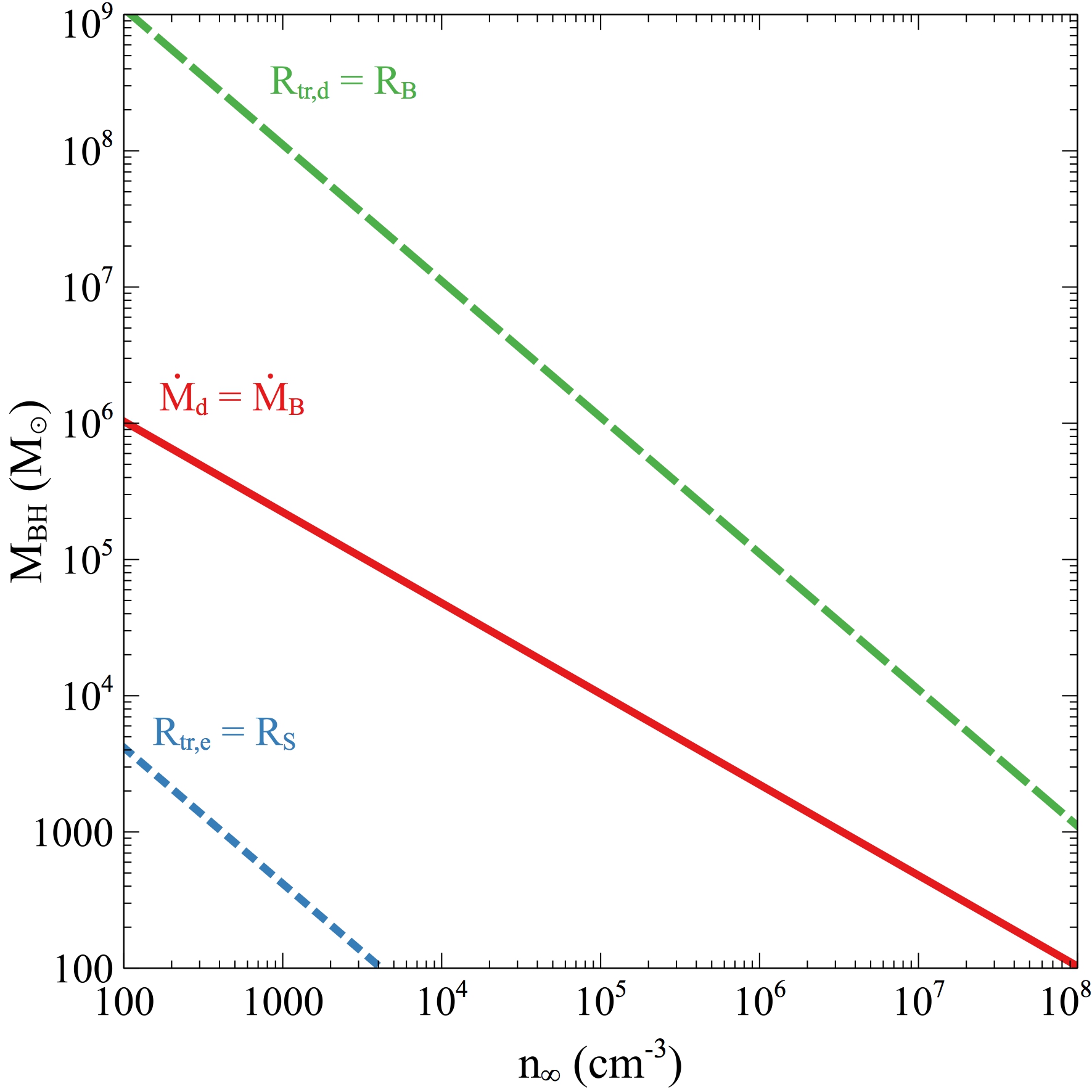}\par
    \includegraphics[width=0.9\linewidth]{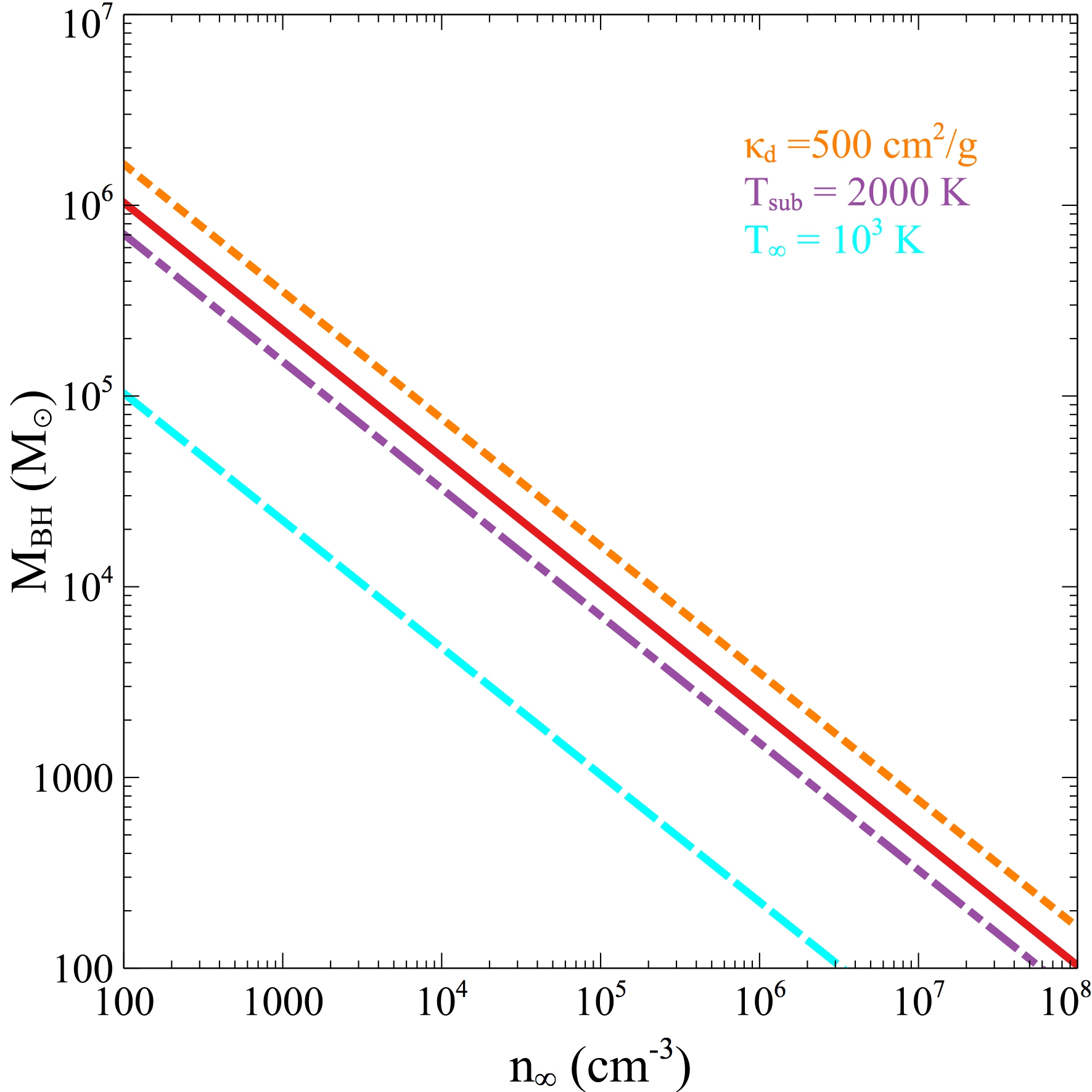}\par 
    \end{multicols}
\caption{ 
The $\mathrm{M_{BH} - n_{\infty}}$ plane with critical boundaries (left panel): $\mathrm{R_{tr,e} = R_S}$ (blue dotted), $\mathrm{\dot{M}_{d} = \dot{M}_B}$ (red solid), and $\mathrm{R_{tr,d} = R_B}$ (green dashed). 
The same $\mathrm{M_{BH} - n_{\infty}}$ plane with variations in the physical parameters (right panel): fiducial parameters (red solid), $T_{\infty} = 10^3$ K (cyan dashed), $\mathrm{T_{sub} = 2000}$ K (violet dash-dot), $\mathrm{\kappa_d = 500 \, cm^2 g^{-1}}$ (orange dotted). 
}
\label{Fig_MBH_ninf_plane}
\end{figure*}


\section{The $M_\textrm{BH} - \MakeLowercase{n_{\infty}}$ plane} 
\label{Section_MBH_n_plane}

Focusing on the two key parameters of the system, namely the central BH mass ($\mathrm{M_{BH}}$) and the host gas density ($n_{\infty}$), a two-dimensional plane can be defined \citep[e.g.][]{Pacucci_et_2017}. 
In our case, the DPT condition $\dot{M}_\textrm{d}/\dot{M}_\textrm{B} < 1$ can be recast into the form 
\begin{align}
&\left( \frac{M_{BH}}{10^5 M_{\odot}} \right)^{3/2} \left( \mathrm{\frac{n_{\infty}}{10^5 cm^{-3}}} \right) \\
\nonumber
&> 0.03 \left( \frac{f_E}{1} \right)^{1/2} \left( \mathrm{\frac{\kappa_d}{10^3 cm^2/g}} \right)^{-1} 
\left( \mathrm{\frac{T_{sub}}{1500 K}} \right)^{-2}  \left( \mathrm{\frac{T_{\infty}}{10^4 K}} \right)^{3/2} 
\label{Eq_MBH_ninf}
\end{align}  

In Figure \ref{Fig_MBH_ninf_plane}, we show the resulting $M_\textrm{BH} - n_{\infty}$ plane. The red solid line defines the DPT condition ($\dot{M}_{d}/\dot{M}_B < 1$): dust photon trapping occurs if the source is located in the region of parameter space above this critical line. The minimal condition for photon trapping due to electron scattering ($\mathrm{R_{tr,e} > R_S}$) is indicated by the blue dotted line. At the other end, the maximal extent of the DPT zone may be estimated e.g. by assuming the Bondi accretion rate in the definition of the trapping radius (equation \ref{Eq_R_trd}). The corresponding maximal DPT radius is smaller than the Bondi radius for fiducial parameters, and here we take $\mathrm{R_{tr,d} \sim R_B}$ as the limiting condition for DPT (green dashed line). On the other side, if the dust photon trapping radius happens to be larger than the Bondi radius, the whole inner region ($r \lesssim R_\mathrm{B}$) could in principle be under trapped conditions, potentially leading to maximally efficient DPT. If we directly impose the physical condition $\mathrm{R_{tr,d} > R_B}$ for a generic mass accretion rate we have $\dot{M} > \frac{4 \pi c}{\kappa_d} \frac{G M_{BH}}{c_s^2}$. Compared to the Bondi rate, the resulting ratio would be $\dot{M}/\dot{M}_B \gtrsim 10$ for fiducial parameters, suggesting that such extreme form of accretion is rather unlikely to be realised. We note that dust photon trapping may be harder to achieve for smaller black hole masses in the early growth phase, but could be facilitated by the higher ambient gas densities at early times (see also Section  \ref{Section_Discussion}). 

Figure \ref{Fig_MBH_ninf_plane} clearly illustrates that a small BH mass requires a high gas density to achieve dust photon trapping, while DPT can occur at lower gas densities for larger BH masses. We further consider how the $M_\textrm{BH} - n_{\infty}$ plane is affected by variations in the other physical parameters (right-hand panel of Figure \ref{Fig_MBH_ninf_plane}). We see that the ambient gas temperature can have a significant impact: for a lower gas temperature, the trapping condition is relaxed and the DPT area becomes larger (the critical line shifts to the lower-left). Similarly, the trapping condition is slightly relaxed for a higher dust sublimation temperature, whereas a lower dust opacity leads to a slight reduction in the DPT area. 

\citet{Pacucci_et_2017} identify a `high-efficiency region' in the $M_\textrm{BH} - n_{\infty}$ plane, where a favourable combination of seed mass and gas density allows efficient accretion and thus rapid BH growth. Requiring that accretion proceeds efficiently all the way from large scales to small scales, they obtain that a black hole with mass $\mathrm{M \gtrsim 10^4 M_{\odot}}$ formed in the high-efficiency region can continuously accrete and grow. Their high-efficiency region is located towards the upper-right of the $M_\textrm{BH} - n_{\infty}$ plane, with a more massive BH being located in the high-efficiency region for a wider range of gas densities. This may roughly correspond to the DPT region represented in Figure \ref{Fig_MBH_ninf_plane}. 

Assuming that photon trapping due to electron scattering limits the emerging luminosity to $\mathrm{L \lesssim L_E}$, 1D RHD simulations indicate that steady hyper-Eddington accretion can occur if the condition $(\mathrm{n_{\infty}/10^5 cm^{-3}}) > (\mathrm{M_{BH}/10^4 M_{\odot}})^{-1} (\mathrm{T_{\infty}/10^4 K})^{3/2}$ is satisfied \citep{Inayoshi_et_2016}. We note that this relation is similar to the DPT condition for $\mathrm{M_{BH} = 10^4 M_{\odot}}$, although the BH mass dependence is steeper in our case. Therefore, in broad agreement with previous studies, we obtain that more massive BHs grow more efficiently, as massive objects can accrete from a wider range of gas densities within the DPT regime. 


\section{Obscured accretion and black hole mass growth}

\subsection{Obscured accretion}
\label{Subsection_Obscured_accretion}

Since photons are trapped within the accretion flow, dust photon trapping necessarily implies obscured accretion. This can have a considerable impact on the observational appearance of the source, such that the early growth of massive black holes may be hidden from view \citep{Wyithe_Loeb_2012, Volonteri_et_2015}. 
A rough estimate of the column density can be obtained by assuming a fraction of the gas free-fall rate at the BH capture radius, as in \citet{Volonteri_et_2015} 
\begin{equation}
\mathrm{N_H} \sim \frac{(\dot{M}/m_p) t_\mathrm{ff}}{r^2} \, , 
\end{equation} 
where $t_\textrm{ff} = \sqrt{\frac{r^3}{2 G M_\mathrm{BH}}}$ is the free-fall timescale, and $r = G M_\mathrm{BH}/\sigma^2$ is the capture radius with velocity dispersion $\sigma$. 
Assuming the critical DPT accretion rate (equation \ref{Eq_Mdotd}), the associated column density is given by
\begin{equation}
\mathrm{N_H} 
\sim \frac{4 \pi c \sigma}{m_p \kappa_d} \sqrt{\frac{c f_E}{2 G M_{BH} \kappa_e \sigma_{SB} T_{sub}^4}} \, . 
\end{equation}

Inserting the numerical values, we obtain
\begin{align}
\mathrm{N_H = 3.5 \times 10^{24} cm^{-2}} &\left( \frac{f_E}{1} \right)^{1/2} \left( \mathrm{\frac{\kappa_d}{10^3 cm^2/g}} \right)^{-1}  \left( \mathrm{\frac{\sigma}{50 km/s}} \right) \\
\nonumber
&\left( \mathrm{\frac{T_{sub}}{1500 K}} \right)^{-2} \left( \frac{M_{BH}}{10^5 M_{\odot}} \right)^{-1/2}
\label{Eq_NH}
\end{align}
For typical parameters, DPT accretion may indeed be obscured, close to the Compton thick limit ($\mathrm{N_H \sim 10^{24} cm^{-2}}$). 
But we note that the column density decreases with increasing BH mass, scaling as $\mathrm{N_H \propto M_{BH}^{-1/2}}$. 
For a massive BH with $\mathrm{M_{BH} \sim 10^9 M_{\odot}}$, the associated column density would be $\mathrm{N_H \sim 7 \times 10^{22} cm^{-2}}$, and the quasar could be unobscured \citep[cf.][]{Volonteri_et_2015}. Such sources may correspond to the luminous quasars detected at $z \gtrsim 6$.  


\subsection{Black hole mass growth}
\label{subsection_BHmass_growth}

As dust photon trapping requires supercritical accretion rates, we expect DPT to be associated with a rapid build-up of the BH mass. 
We recall that the standard Eddington accretion rate can be written as
\begin{equation}
\dot{M} = \frac{1 - \epsilon}{\epsilon} \frac{M}{t_E} \, ,
\label{Eq_dotME}
\end{equation}
where $t_E = \frac{\sigma_T c}{4 \pi G m_p}$ is the Eddington timescale. 
Integrating equation (\ref{Eq_dotME}), we obtain the corresponding BH mass growth 
\begin{equation}
M(t) = M_0 \exp \left( \frac{1-\epsilon}{\epsilon} \frac{t}{t_E} \right) \, , 
\end{equation}
where $M_0$ is the initial seed mass. 

In a similar way, integrating the DPT accretion rate (equation \ref{Eq_Mdotd}), we obtain
\begin{equation}
M(t) 
= \left( \sqrt{M_0} + \frac{2 \pi c}{\kappa_d} \sqrt{ \frac{G c f_E}{\kappa_e \sigma_{SB} T_{sub}^4}} t \right)^2 \, . 
\label{Eq_MBH_DPT}
\end{equation} 

Let us consider a simple numerical example of a seed BH accreting for a Salpeter time ($t \cong 5 \times 10^7$ yr). Starting from the same initial mass of $\mathrm{M_0 \sim 10^5 M_{\odot}}$, the final mass is $\mathrm{M_{BH} \sim 3 \times 10^5 M_{\odot}}$ for the Eddington accretion rate, while it is $\mathrm{M_{BH} \sim 7 \times 10^8 M_{\odot}}$ for the DPT accretion rate. In contrast to the case of Eddington-limited accretion, a BH can grow considerably under DPT conditions. Therefore DPT accretion may enable the growing BHs to reach large masses within short timescales, and may help explain the existence of supermassive BHs in the early Universe. 

The DPT accretion rate (equation \ref{Eq_Mdotd}) is the minimal accretion rate required to achieve dust photon trapping; so the associated mass growth may be considered as a lower limit to the actual BH growth. However, in the presence of dust grains, one should also consider the effective Eddington limit for dusty gas \citep[e.g.][]{Fabian_2012}, which is lower than the standard Eddington luminosity. A more detailed analysis will be required to investigate the interplay between radiation pressure on dust and dust photon trapping. 
 

\section{Discussion}
\label{Section_Discussion}

Recent observations show that the host galaxies of the first quasars at $z \gtrsim 6$ are chemically evolved systems, containing unexpectedly large amounts of dust and metals \citep[e.g.][and references therein]{Valiante_et_2017}. ALMA observations of a sample of quasars at $z \gtrsim 6$ indicate dust masses ranging from $\mathrm{M_d \sim 10^7 M_{\odot}}$ for the faintest sources to $\mathrm{M_d \sim 10^9 M_{\odot}}$ for the brightest objects \citep{Venemans_et_2018}. 
A large dust mass of $\mathrm{M_d \sim 7 \times 10^7 M_{\odot}}$ is even observed in the highest-redshift quasar at $z = 7.6$ \citep{Wang_et_2021}. The presence of dust grains at such high redshifts requires rapid dust formation, possibly due to early supernovae and grain growth in the interstellar medium \citep{Lesniewska_Michalowski_2019}. 
The above observational results point towards fast dust enrichment occurring in parallel to the early growth of massive black holes within the first Gyr of the Universe. This prompts us to consider the potential role of dust photon trapping in the supercritical growth of massive BHs at early times. 


\subsection{Physical implications}
\label{subsection_physical_implications}

Photon trapping affects the radiative luminosity output, and thus can influence the accretion flow onto the growing BH, as well as determine the observational appearance of the source. Due to photon trapping, the emergent diffusive luminosity is usually limited to values lower than the Eddington luminosity, independently of the accretion rate. This can be seen from the simple relation $\mathrm{L \lesssim G M_\mathrm{BH} \dot{M}/R_\mathrm{tr} = L_E}$, whereby the accretion rate terms cancel out (as $R_\mathrm{{tr}} \propto \dot{M}$). Because the emergent luminosity does not exceed the Eddington limit, radiation pressure can not prevent the gas inflow, and the resulting accretion rates can in principle reach arbitrarily high values \citep{Begelman_1979}. This provides a physical mechanism potentially allowing super-Eddington accretion. Indeed, as the photons are trapped within the accretion flow, radiation pressure feedback becomes inefficient, such that the accretion rates can reach values far exceeding the Eddington rate \citep{Wyithe_Loeb_2012}.

When the emergent luminosity is limited by photon trapping ($\mathrm{L \lesssim L_E}$), the size of the ionised HII region is also likely confined inside the Bondi radius ($\mathrm{R_{HII} < R_B}$). As a consequence, photoionisation feedback can not operate efficiently and is unable to halt the inflowing gas, potentially leading to hyper-Eddington accretion \citep{Inayoshi_et_2016}. Moreover, the effect of photoionisation feedback may be further reduced for realistic accretion disc spectra, compared to the case of the single power-law spectrum usually assumed in most works \citep{Takeo_et_2019}. 

As mentioned in Section \ref{Section_DPT_parameterspace}, for accretion rates satisfying the DPT condition, photon trapping due to electron scattering is also naturally realised. In the case of dust photon trapping, radiation feedback at radii larger than the dust sublimation radius is effectively suppressed, while the entire accretion disc radiation may be efficiently trapped. Under such extreme trapping conditions, most of the infalling matter may be accreted \citep{Volonteri_et_2015}. Since the accretion rate leading to dust photon trapping is much larger than the Eddington rate, it should be associated with super-Eddington accretion and a rapid build-up of the BH mass. For instance, adopting the DPT accretion rate as the minimal value, the enhancement factor could be of a few orders of magnitude compared to the Eddington-limited value. 

It is known that photon trapping also modifies the emergent spectral energy distribution. Higher energy photons originating in the inner accretion flow are more efficiently trapped than lower energy photons produced from the outer regions.  As a result, the peak frequency of the spectral energy distribution shifts towards lower energies, and the AGN spectrum becomes softer with increasing accretion rate \citep{Ohsuga_et_2003}. In the case of DPT,  the whole disc radiation can be efficiently trapped, and the overall luminosity output may be considerably reduced, leading to faint AGNs. This may not be the case for photon trapping due to electron scattering, where only the inner parts of the accretion disc remain trapped (unless the accretion rate is extremely high). 

Since UV/optical photons are trapped within the dusty accretion flow, the central AGN will be obscured. In this context, X-ray photons would be even more readily trapped than UV/optical photons, and thus super-Eddington accretion will be likely obscured at both X-ray and UV/optical wavelengths \citep{Wyithe_Loeb_2012}. In our picture, DPT accretion is indeed associated with significant obscuration, with large column densities expected at early times. Cosmological hydrodynamic simulations also suggest heavy obscuration on different physical scales around the growing black holes in the early Universe \citep{Trebitsch_et_2019, Ni_et_2020}. Assuming a standard radiative efficiency of $\epsilon \sim 0.1$, it has been estimated that the obscured fraction at $z > 7$ could be of the order of $\gtrsim 80\%$ \citep{Davies_et_2019}. In fact, much of the early growth of massive black holes could be obscured, and a large number of such obscured quasars may be missed in current optical surveys \citep[][and references therein]{Hickox_Alexander_2018, Inayoshi_et_2020}. 

On the other hand, obscuration by dusty gas may cause higher energy photons to be reprocessed into lower energy photons, leading to an increase in the infrared (IR) emission. In high column density sources (e.g. Compton-thick AGNs), the X-ray emission will be significantly suppressed, while the emission in the IR band could be considerably enhanced. Future instruments may allow the detection of such obscured accreting BHs at very high redshifts \citep[][and references therein]{Pacucci_et_2019}.


\subsection{Comparison to other works}

Different forms of supercritical accretion have been proposed and discussed in the literature (see Introduction). While most previous works consider primordial gas composition, we discuss below a few other studies that explicitly analyse the accretion of dusty gas. 

\citet{Wyithe_Loeb_2012} consider photon trapping in dusty accretion flows, and discuss the cosmological conditions allowing super-Eddington accretion. We note that their minimal accretion rate required for photon trapping is essentially equivalent to our DPT condition (equation \ref{Eq_Mdotd}). But they define the sublimation radius as a fixed multiple of the gravitational radius, while we assume a physically motivated sublimation radius, explicitly depending on the underlying parameters (Section \ref{Subsection_photon_trapping}). 
The DPT condition may actually evolve with cosmic time, due to the redshift-dependence of the ambient gas density. While the number density is constrained by the halo properties in \citet{Wyithe_Loeb_2012}, here we leave $n_{\infty}$ as a free parameter and explore a wide range of ambient gas densities \citep[see also][]{Pacucci_et_2017}. Qualitatively, higher gas densities are to be expected at higher redshifts. As a consequence, when the ambient gas density was higher in the past, dust photon trapping could occur even for low-mass BHs, enabling efficient growth at early times. This suggests that the physical conditions in the early Universe were most favourable for the occurrence of DPT, allowing rapid BH growth at critical times. 
 
 \citet{Yajima_et_2017} perform 1D RHD simulations of dusty gas accretion in the low density regime ($n_{\infty} \sim 10-10^3 \, \mathrm{cm^{-3}}$) to analyse the interplay between dust and photoionisation feedback. Compared to the accretion of primordial gas characterised by strong periodic bursts, the accretion pattern of dusty gas is found to be much smoother with smaller fluctuations. Radiation pressure on dust (neglecting the reprocessed IR emission) seems to regulate the overall accretion flow, while the effect of dust opacity (in reducing the size of the HII region) seems to be of secondary importance. In our framework, we would not expect efficient dust photon trapping to occur in such low density environments, unless the central BH is already quite massive (Section \ref{Section_MBH_n_plane}). 

Extending the investigation to a higher density regime ($n_{\infty} \sim 10^5 \mathrm{cm^{-3}}$), \citet{Toyouchi_et_2019} carry out 1D RHD simulations of dusty gas to analyse the possible transition to hyper-Eddington accretion. By explicitly considering the radiative force due to reprocessed IR photons, which can penetrate far into the neutral medium beyond the HII bubble, the resulting accretion rates are reduced compared to the Bondi value. Moreover, the accretion rates are found to be lower for higher metallicities, suggesting that super-Eddington accretion is prevented above a certain critical metallicity. However, an opposite result is obtained in the newest 3D RHD simulations including radiation anisotropy and mass outflows, whereby higher metallicities lead to higher accretion rates \citep{Toyouchi_et_2020}. 

In this work, we just consider a characteristic dust opacity at UV wavelengths ($\kappa_\mathrm{d,UV} \sim 10^3 \, \mathrm{cm^2} \mathrm{g^{-1}}$).  UV photons are most efficiently absorbed by dust grains, and subsequently reprocessed as IR photons (by energy conservation). The dust opacities in the IR are much lower than in the UV band, with a typical value of $\kappa_\mathrm{d,IR} \sim 5 \, \mathrm{cm^2}\mathrm{g^{-1}}$ scaling with the dust-to-gas ratio. As a consequence, the associated trapping radius is smaller by a factor $\sim (\kappa_\mathrm{d,IR}/\kappa_\mathrm{d,UV})$, and the critical accretion rate for IR photon trapping is correspondingly larger. Therefore the condition for the trapping of reprocessed IR radiation is likely more stringent, and would require accretion rates in excess of the Bondi rate. A more detailed analysis of the impact of such IR photon trapping is left for future work. In reality, the actual dust opacity also depends on grain size, structure, and composition, as well as incident radiation spectrum. Since the dust opacity grows in proportion to the dust-to-gas ratio, which in turn may scale with the metallicity, the critical DPT accretion rate is expected to be lower for higher metallicities. This suggests that dust photon trapping could be more easily realised in higher metallicity environments.

In addition, supercritical accretion can be facilitated by anisotropic feedback, either due to anisotropic emission from the accretion disc or shadowing effects on larger scales \citep{Sugimura_et_2017, Takeo_et_2018}. The exact radiation pattern is ultimately determined by the central BH spin, with low/high spins leading to prolate/oblate emission patterns, respectively \citep{Ishibashi_et_2019}. In reality, several competing mechanisms operate simultaneously, leading to opposing effects. For instance, the inclusion of the gas angular momentum and associated centrifugal barrier can hinder supercritical accretion \citep{Sugimura_et_2018}, but an accretion disc geometry may also favour the escape of IR photons in the polar directions, reducing the overall radiative feedback \citep{Toyouchi_et_2019}. Contrary to what might be expected, it has been shown that the transition to the hyper-Eddington accretion regime could even be facilitated by powerful outflows \citep{Takeo_et_2020}. 

We have previously discussed how AGN radiative feedback, driven by radiation pressure on dust, may affect the surrounding environment through the development of galactic outflows \citep{Ishibashi_Fabian_2015, Ishibashi_et_2018}. We have also considered the effects of such radiation feedback-driven outflows on the early growth of massive BHs in the case of Eddington-limited accretion \citep{Ishibashi_2019}. A further step will be to investigate the dynamical coupling between radiative feedback, dust photon trapping, and supercritical BH growth in the early Universe.


\section*{Acknowledgements }

WI acknowledges support from the University of Zurich.


\section*{Data availability}

No new data were generated or analysed in support of this research.

  
\bibliographystyle{mn2e}
\bibliography{biblio.bib}


\label{lastpage}

\end{document}